\documentstyle[prd,aps,floats,amsfonts,amssymb,amsmath]{revtex}
\begin{document}
 \draft
 \title{Discrete symmetries and the muon's gyro-gravitational ratio in $g-2$ experiments}
\author{G. Lambiase$^{a,b}$\thanks{E-mail: lambiase@sa.infn.it}, G. Papini$^{c,d}$\thanks{E-mail:
            papini@uregina.ca}}
\address{$^a$Dipartimento di Fisica "E.R. Caianiello"
 Universit\'a di Salerno, 84081 Baronissi (Sa), Italy.}
  \address{$^b$INFN - Gruppo Collegato di Salerno, Italy.}
  \address{$^c$Department of Physics, University of Regina, Regina, Sask, S4S0A2, Canada.}
  \address{$^d$International Institute for Advanced Scientific Studies, 89019 Vietri sul Mare (SA), Italy.}
\date{\today}
\maketitle
\begin{abstract}
We show that recent, persistent discrepancies between theory and
experiment can be interpreted as corrections to the
gyro-gravitational ratio of the muon and lead to improved upper
limits on the violation of discrete symmetries in rotational
inertia.
\end{abstract}
\pacs{PACS No.: 11.30.Er, 04.20.Cv, 04.80.-y, 13.40.Em}

In a recent comment one of us (G.P.) suggested that discrepancies
between the experimental and standard model values of the muon's
anomalous magnetic moment $ a_{\mu}(exp)-a_{\mu}(SM)$ lead to
upper limits on the violation of discrete symmetries in the
spin-rotation coupling \cite{papini2}. Improved values of $
a_{\mu}(exp)$ for positive and negative muons and persistent,
residual differences $ a_{\mu}(exp)-a_{\mu}(SM)$ can now be used
to reduce the extent of $ P$ and $ T$ violation in spin-rotation
coupling and also to determine the parameters $ \epsilon_{+}$ and
$ \epsilon_{-}$ of \cite{papini2}. These quantities can be
interpreted as corrections to the gyro-gravitational ratio of the
muon. The argument is summarized below.

$g-2$ experiments involve muons in storage rings \cite{farley}. As
the muons decay, the angular distribution of those electrons
projected forward in the direction of motion reflect the
precession of the muon spin along the cyclotron orbits. The value
of $ a_{\mu}$ is determined experimentally from the number of
decay electrons or positrons \cite{brown3,brown4}
\begin{equation}\label{1bis}
N\left(t\right)= N_{0}e^{-t/\left(\gamma\tau\right)}\left[1+ C
\cos \left(a_{\mu}\frac{eBt}{m}+
 \phi_{a} \right)\right],
\end{equation}
where $ m $ is the mass of the muon and $ \gamma\tau $ its dilated
lifetime. $ N_{0}, C$ and $ \phi_{a}$ depend on the energy
threshold selected. Equation (\ref{1bis}) can be directly related
to the muon Hamiltonian $ H $ that follows from the Dirac equation
in the muon's rotating frame. $ H$ can be split into a part $
H_{0}$ that is diagonal and contributes only to the overall energy
$ E $ of the states, and a part
\begin{equation}\label{1ter}
H'=\left(-\frac{\hbar}{2}\vec{\omega}-\mu\vec{B}\right)\cdot\vec{\sigma}
\end{equation}
that accounts for spin precession. The spin-rotation coupling in
(\ref{1ter}) is also known as the Mashhoon term \cite{MASH}.
$\displaystyle{\mu=\left(1+a_{\mu}\right)\mu_0}$ represents the
total magnetic moment of the muon. Before decay the muon states
can be represented as
\begin{equation}\label{2}
  |\psi(t)>=a(t)|\psi_+>+b(t)|\psi_->\,,
\end{equation}
where $|\psi_+>$ and $|\psi_->$ are the right and left helicity
states of the Hamiltonian $H_0$. For simplicity all quantities are
taken to be time-independent. It is also assumed that the effects
of electric fields used to stabilize the orbits and stray radial
electric fields are cancelled by choosing an appropriate muon
momentum \cite{farley}. The total Hamiltonian $H=H_0+H'$ is most
conveniently referred to a left-handed triad of axes rotating
about the $x_2$-axis in the direction of motion of the muons. The
$x_3$-axis is tangent to the orbits and in the direction of the
muon momentum. The magnetic field is $B_2=-B$. As shown in
\cite{papini1}, $ H'$ by itself accounts for the main features of
(\ref{1bis}).

Assume now that the coupling of rotation to $\mid\psi_{+}>$
differs in strength from that to $\mid\psi_{-}>$ as in
\cite{papini2}. This can be accomplished by multiplying the
Mashhoon term in (\ref{1ter}) by the matrix
$A=\left(\begin{array}{cc}\kappa_{+}& 0 \cr 0 &
\kappa_{-}\end{array}\right)$ that reflects the different coupling
strength of rotation to the two helicity states.
 A violation of $P$ and $T$ in $ H $ thus arises through
$\kappa_{+}-\kappa_{-}\neq 0$. The constants $\kappa_{+}$ and
$\kappa_{-}$ are assumed to differ from unity by small amounts
$\epsilon_{+}$ and $\epsilon_{-}$. Persistent, residual
discrepancies in very high precision measurements of  $ a_{\mu}$
for both positive and negative muons \cite{brown3,brown4} can now
be used to find the values of both $ \epsilon_{+}$ and $
\epsilon_{-}$. The same discrepancies yield upper limits on $P$
and $T$ invariance violations in spin-rotation coupling.

The coefficients $a(t)$ and $b(t)$ in (\ref{2}) evolve in time
according to
\begin{equation}\label{5a}
i\frac{\partial}{\partial t}
            \left(\begin{array}{c}
            a(t) \cr b(t)\end{array}\right)=
            M\left(\begin{array}{c}
            a(t) \cr b(t)\end{array} \right),
\end{equation}
where
\begin{equation}\label{6b}
M=\left(\begin{array}{cc}
  E-i\frac{\Gamma}{2}& i\left(\kappa_{+}\frac{\omega_{2}}{2}-\mu B\right)\cr
 -i\left(\kappa_{-}\frac{\omega_{2}}{2}-\mu B\right)& E-i\frac{\Gamma}{2}\end{array}\right),
\end{equation}
and $\Gamma$ represents the width of the muon. The spin-rotation
term is off-diagonal in (\ref{6b}) and does not therefore couple
to matter universally. It violates Hermiticity as shown in
\cite{papini2} and, in a general way, in \cite{scolarici}. It also
violates $T$, $P$ and $PT$, while nothing can be said about $CPT$
conservation which requires $H$ to be Hermitian
\cite{kennysachs,sachs}. Non-Hermitian corrections to the width of
the muon are of second order in $\epsilon_{\pm}$'s and are
neglected.

$M$ has eigenvalues $ h = E - i \Gamma /2 \pm R $ and eigenstates
\begin{eqnarray}
|\psi_{1}>&=& N\left[\eta|\psi_{+}>+|\psi_{-}>\right],\nonumber \\
|\psi_{2}>&=& N\left[-\eta|\psi_{+}>+|\psi_{-}>\right],
\end{eqnarray}
where
\begin{equation}
R=\sqrt{\left(\kappa_{+}\frac{\omega_{2}}{2}-\mu
B\right)\left(\kappa_{-}\frac{\omega_{2}}{2}-\mu B\right)},
\end{equation}
$|N|^{2}= 1/\left(1+|\eta|^{2}\right)$ and $
\eta=\frac{i}{R}\left(\kappa_+\frac{\omega_{2}}{2}-\mu B\right)$.
Then the muon states (\ref{2}) that satisfy the condition
$|\psi(0)>=|\psi_{-}>$ are
\begin{equation}
|\psi(t)>=\frac{e^{-iEt-\frac{\Gamma t}{2}}}{2}[-2i\eta \sin Rt
|\psi_{+}>+2\cos Rt |\psi_{-}>],
\end{equation}
and the spin-flip probability is
\begin{eqnarray}\label{9}
P_{\psi_{-}\rightarrow
 \psi_{+}}&=&|<\psi_{+}|\psi(t)>|^{2}\nonumber \\
          &=&\frac{e^{-\Gamma t}}{2}
          \frac{\kappa_{+}\omega_{2}-2\mu B}{\kappa_-\omega_{2}-2\mu B}
          \left[1-\cos \left(2Rt\right)\right].
\end{eqnarray}
When $ \kappa_{+}=\kappa_{-}=1$, (\ref{9}) reproduces the
essential features of  (\ref{1bis})\cite{papini1}.

Substituting $\kappa_{+}=1+\epsilon_{+},
 \kappa_{-}=1+\epsilon_{-}$ into (\ref{9}), one finds
\begin{equation}\label{11}
 P_{\psi_{-}\rightarrow\psi_{+}}=\frac{e^{-\Gamma t}}{2}
 \frac{\epsilon_{+}-a_{\mu}}{\epsilon_{-}-a_{\mu}}
 \left[1-\cos\left(2Rt\right)\right].
\end{equation}
A similar expression for $ P_{\psi_{+}\rightarrow\psi_{-}}$ can be
obtained starting from the condition $|\psi(0)>=|\psi_{+}>$.

We attribute the discrepancy between $a_{\mu}(exp)$ and
$a_{\mu}(SM)$ to a violation of the conservation of the discrete
symmetries by the spin-rotation coupling term $
-\frac{1}{2}A\omega_{2}\sigma^{2}$. The upper limit on the
violation of $P,T$ and $PT$ is derived from (\ref{11}) assuming
that the deviation from the current value of $a_{\mu}(SM)$ is
wholly due to $\epsilon_{\pm}$. The most precise sets of data yet
give $a_{\mu_{+}}(exp)-a_{\mu_{+}}(SM)\equiv b=26\times 10^{-10}$
for positive muons \cite{brown3} and
$a_{\mu_{-}}(exp)-a_{\mu_{-}}(SM)\equiv d=33\times 10^{-10}$ for
negative muons \cite{brown4}. These then are the  upper limits to
the violation of the discrete symmetries.

At the same time the two values of $a_{\mu}(exp)-a_{\mu}(SM)$ are
due, in the model, to the different coupling strengths between
rotational inertia and the two helicity states of the muon. The
values of $ \epsilon_{+}$ and $ \epsilon_{-}$ can be determined
from $ \cos\left(2 R t\right)$ in (\ref{9}). The equations are
\begin{equation}\label{11a}
\left(a_{\mu_{+}}-
\epsilon_{+}\right)\left(a_{\mu_{+}}-\epsilon_{-}\right)=b^{2}
\end{equation}
and
\begin{equation}\label{11b}
\left(a_{\mu_{-}}-
\epsilon_{+}\right)\left(a_{\mu_{-}}-\epsilon_{-}\right)=d^{2}.
\end{equation}
The reality condition that follows from the solutions of
(\ref{11a}) and (\ref{11b}) is satisfied, for $ b$ and $ d$ fixed,
by ranges of values of $ a_{+}$ and $ a_{-}$ that are compatible
with present experimental accuracies.
 Equations (\ref{11a}) and (\ref{11b}) have the approximate
solutions
\begin{equation}\label{11c}
\epsilon_{+}\simeq \frac{a_{\mu_{+}}+
a_{\mu_{-}}}{2}-\frac{d^{2}-b^{2}}{2\left(a_{\mu_{-}}-a_{\mu_{+}}\right)}
\end{equation}
and
\begin{equation}\label{11c1}
\epsilon_{-}\simeq a_{\mu_{+}}+ \frac{2
b^{2}\left(a_{\mu_{-}}-a_{\mu_{+}}\right)}{\left(a_{\mu_{-}}-a_{\mu_{+}}\right)^{2}+\left(d^{2}-b^{2}\right)}.
\end{equation}
More precise, numerical solutions give $ \epsilon_{+}\simeq
11659189 \cdot 10^{-10},  \epsilon_{-} \simeq 11659152 \cdot
10^{-10}$ and $ \Delta\epsilon \equiv
\epsilon_{+}-\epsilon_{-}\simeq 37.65878 \cdot 10^{-10}$. These
values are significant in view of the precision with which $
a_{\mu\pm}, b, d$ have been determined. In our simple model,
therefore, the coupling of rotation to positive helicity is larger
than that to negative helicity.

In conclusion, muons in storage rings are rotating quantum
gyroscopes that are sensitive probes of rotational inertia.
Extremely precise $ g-2$ experiments also concern the violation of
the equivalence principle in quantum mechanics \cite{LA,MA} and
the conservation of discrete symmetries. Possibly, the deviations
$a_{\mu}(exp)-a_{\mu}(SM)$ will be ultimately explained in ways
more conventional than inertial anomalies. However, the upper
limits on the violations of $ P$ and $ T$ in inertia-gravity
interactions that can be reached by $ g-2$ experiments are at
least as sensitive as those obtained by other means
\cite{schiff,leitner,dass,almeida}. There are some important
differences too. The $g-2$ measurements are performed in strictly
controlled laboratory conditions. In addition, the limits obtained
apply to terms in $ H'$ that make themselves manifest in an
essential way in $ g-2$ experiments and are an inescapable
consequence of the covariant Dirac equation. The validity of this
equation in an inertial-gravitational context finds support in the
experimental verifications of the Page-Werner \cite{page} and
Bonse-Wroblewski \cite{bonse} effects.

If the deviations $a_{\mu}(exp)-a_{\mu}(SM)$ are entirely due to a
difference in the coupling strength between rotation and the
helicity states, then the gyro-gravitational ratio of the muon is
no longer unity and the violation of $ P$ and $ T$ is relatively
stronger for positive helicity. This also means that rotating
objects are inherent sources of $ P$ and $T$ violation.

In derivations based on the covariant Dirac equation, the coupling
of inertia and gravitation to spin is identical to that for
orbital angular momentum. It is then said that the
gyro-gravitational ratio of a spin$-1/2$ particle is one
\cite{DE,audr,kann,singh}. A suggestive interpretation of this
result is that the internal distributions of the gravitational
mass, associated with the interaction, and of inertial mass,
associated with the angular momentum, equal each other \cite{DE}.
This is no longer so when $ \epsilon_{\pm}\neq 0$. There is a
certain similarity, here, with the electromagnetic case where $
g=2 $ is required by the Dirac equation, but not by quantum
electrodynamics. The deviations of $ \kappa_{+} $ and $
\kappa_{-}$ from unity that are consistent with $ g-2$ experiments
are both of the order of $ a_{\mu}$, or $ \simeq 10^{-3}$, and
differ from each other by $ \Delta\epsilon \simeq 3.7\cdot
10^{-9}$. While small values of $ \epsilon_{\pm} $ do not give
rise to measurable mass differences in macroscopic objects
\cite{papini2}, violations of the discrete symmetries may have
interesting astrophysical and cosmological implications.


\begin{thebibliography}{00}
\bibitem{papini2} G. Papini, Phys. Rev. D  {\bf 65}, 077901 (2001).
\bibitem{farley} F.J.M. Farley and E. Picasso in \emph{Advanced Series in
        High-Energy Physics, vol. 7, Quantum Electrodynamics}, edited by T.
        Kinoshita (World Scientific, Singapore 1990), p. 479.
\bibitem{brown3} G. W Bennett $et$ $al.$, Muon (g-2) Collaboration, Phys. Rev. Lett. {\bf
                 89}, 101804 (2002); Phys. Rev. Lett. {\bf 92}, 161802 (2004), arXiv:hep-ex/0401008 v3 21
                 Feb 2004.
\bibitem{brown4} H. N. Brown $et$ $al.$, Muon (g-2) Collaboration,
                 Phys. Rev. D {\bf 62}, 091101 (2002); Phys. Rev.
                 Lett. {\bf 86}, 2227 (2001).
\bibitem{MASH} B. Mashhoon, Phys. Rev. Lett. {\bf 61}, 2639 (1988).
\bibitem{papini1} G. Papini, in \emph{Advances in the interplay between quantum and gravity physics},
                edited by Peter G. Bergmann and V. de Sabbata (Kluwer Academic,
                Dordrecht 2002) p.317 , arXiv:gr-qc/0110056. \\
                 G. Papini, G. Lambiase, Phys. Lett. A {\bf294}, 175 (2002).
\bibitem{scolarici} G. Scolarici, L. Solombrino, Phys. Lett. A {\bf 303}, 239 (2002).
\bibitem{kennysachs} Brian G. Kenny and Robert G. Sachs, Phys. Rev. D {\bf8}, 1605 (1973).
\bibitem{sachs} Robert G. Sachs, Phys. Rev. D {\bf33}, 3283 (1986).
\bibitem{LA} C. L\"{a}mmerzahl, General Relativity and Gravitation {\bf 28}, 1043 (1996).
\bibitem{MA} Ph. D. Mannheim in \emph{Contemporary Fundamental Problems}, edited by Valeri Dvoeglazov
            (Nova Science Publishers, New York 2000), arXiv:gr-qc/9810087.
\bibitem{schiff} L.I. Schiff, Phys. Rev. Lett. {\bf1}, 254 (1958).
\bibitem{leitner} J. Leitner and S. Okubo, Phys. Rev. {\bf136}, B1542 (1964).
\bibitem{dass} N.D. Hari Dass, Phys. Rev. Lett. {\bf36}, 393 (1976); Ann. Phys.(NY)
                {\bf107}, 337 (1977).
\bibitem{almeida} L.D. Almeida, G.E.A. Matsas and A.A. Natale, Phys. Rev. D {\bf39}, 677 (1989).
\bibitem{page} S. A. Werner, J.-L. Staudenmann and R. Colella,
               Phys. Rev. Lett. {\bf 42}, 1103 (1979). L. A. Page, Phys. Rev.
               Lett. {\bf 35}, 543 (1975).
\bibitem{bonse} U. Bonse and T. Wroblewski, Phys. Rev. Lett. {\bf 51}, 1401 (1983).
\bibitem{DE} C. G. De Oliveira and J. Tiomno, Nuovo Cimento
               {\bf 24}, 672 (1962).
\bibitem{audr} J. Audretsch, J. Phys A: Math. Gen. {\bf 14}, 411 (1981).
\bibitem{kann} L. Kannenberg, Ann. Phys. (N.Y.) {\bf 103}, 64 (1977).
\bibitem{singh} D. Singh and G. Papini, Nuovo Cimento B {\bf
                 115}, 223 (2000).
\end{thebibliography}
\end{document}